\documentclass[letter,twocolumn]{jpsj2} 
\usepackage{psfrag}
\newcommand{\nn}{\nonumber}
\newcommand{\bra}{\langle}
\newcommand{\ve}{\vert}
\newcommand{\ket}{\rangle}

\title{On Which Length Scales Can Temperature Exist in Quantum Systems?}

\author{\textsc{Michael Hartmann}$^{1,2}$\thanks{E-mail address: michael.hartmann@dlr.de}, \textsc{G\"unter Mahler}$^{2}$ and \textsc{Ortwin Hess}$^{3}$}

\inst{$^{1}$Institute of Technical Physics, DLR-Stuttgart, Pfaffenwaldring 38-40, 70569 Stuttgart, Germany \\
$^{2}$Institute of Theoretical Physics I, University of Stuttgart, Pfaffenwaldring 57, 70550 Stuttgart, Germany \\
$^{3}$ School of Electronics \& Physical Sciences, University of Surrey, Guildford, GU2 7XH, United Kingdom}

\abst{We consider a regular chain of elementary quantum systems with nearest neighbor
interactions and assume that the total system is in a
canonical state with temperature $T$. We analyze under what condition the state factors
into a product of canonical density matrices with respect to groups of $n$ subsystems
each, and when these groups have the same temperature $T$. While in classical mechanics
the validity of this procedure only depends on the size of the groups $n$, in quantum
mechanics the minimum group size $n_{\text{min}}$ also depends on the temperature $T \,$!
As examples, we apply our analysis to different types of Heisenberg spin chains.}

\kword{quantum many particle systems, local temperature, correlations}

\begin{document}
\maketitle

Thermodynamics is among the most successfully and extensively applied theoretical concepts in
physics. Notwithstanding, the various limits of its applicability are not fully understood
\cite{Allahverdyan2000,Hill2001,Rajagopal2004}.

Of particular interest is its microscopic limit.
Down to which length scales can its standard concepts meaningfully be defined and employed?

Besides its general importance, this question has become increasingly relevant recently since
amazing progress in the synthesis and processing of materials with structures on
nanometer length scales has created a demand for better understanding of thermal properties of
nanoscale devices, individual nanostructures and nanostructured materials
\cite{Cahill2003}.
Experimental techniques have improved to such an extent that the measurement of thermodynamic
quantities like temperature with a spatial resolution on the nanometer scale seems within reach
\cite{Gao2002,Pothier1997}. 

To provide a basis for the interpretation of present day and future experiments in nanoscale physics
and technology and to obtain a better understanding of the limits of thermodynamics,
it is thus indispensable to clarify the applicability of thermodynamical concepts
on small length scales starting from the most fundamental theory at hand, i. e. quantum mechanics.
In this context, one question appears to be particularly important and interesting:
Can temperature be meaningfully defined on nanometer length scales?

The existence of thermodynamical quantities, i. e. the existence of the thermodynamical limit strongly
depends on the correlations between the considered parts of a system \cite{Saito1996}.
With increasing size, the volume of a region in space grows faster than its surface.
Thus effective interactions between two regions, provided they are short ranged, become less relevant
as the sizes of the regions increase \cite{Ruelle1969,Hartmann2003a}.
The scaling of interactions between parts of a system compared to the energy contained in the parts
themselves thus sets a minimal length scale on which correlations are
still small enough to permit the definition of local temperatures.
It is the aim of this letter to study this connection quantitatively

For this purpose, the applied definition of temperature becomes crucial.
We define local temperature to exist if the respective part of the system is in a
canonical state. Besides its motivation from statistical mechanics, there are also rather practical
arguments for this definition. 

The canonical distribution is an exponentially decaying function of energy characterized by one
single parameter \cite{Kubo1985}. This implies that there is a one-to-one mapping between temperature and
the expectation values of observables, by which temperature is usually measured \cite{Hartmann2004c}.

Furthermore, the product of this distribution times the density of states, which is typically
a strongly growing function of energy \cite{Tolman1967} for large systems,
forms a pronounced peak and thus physical quantities like energy have ``sharp'' values.

Based on the above arguments and noting that a quantum description becomes imperative at nanoscopic scales,
the following approach appears to be reasonable:
Consider a large homogeneous quantum system, brought into a thermal state via interaction with its environment,
divide this system into subgroups and analyze for what
subgroup-size the concept of temperature is still applicable.

Recently, spin chains have been subject of extensive studies in condensed matter physics.
We therefore study various types of Heisenberg chains with respect to our present purpose.

We consider a homogeneous (i.e. translation invariant) chain of elementary quantum
subsystems with nearest neighbor interactions.
The Hamiltonian of our system is thus of the form
\begin{equation}\label{hamil}
H = \sum_{i} H_i + I_{i,i+1},
\end{equation}
where the index $i$ labels the elementary subsystems. $H_i$ is the Hamiltonian of subsystem $i$
and $I_{i,i+1}$ the interaction between subsystem $i$ and $i+1$.
We assume periodic boundary conditions.

We now form $N_G$ groups of $n$ subsystems each
(index $i \rightarrow (\mu-1) n + j; \: \mu = 1, \dots, N_G; \: j = 1, \dots, n$)
and split this Hamiltonian into two parts,
\begin{equation}
\label{hsep}
H = H_0 + I,
\end{equation}
where $H_0$ is the sum of the Hamiltonians of the isolated groups,
\begin{eqnarray}\label{isogroups}
H_0 & = & \sum_{\mu=1}^{N_G} \left( \mathcal{H}_{\mu} - I_{\mu n,\mu n + 1} \right) \enspace \enspace
\text{with} \nn \\
\mathcal{H}_{\mu} & = & \sum_{j=1}^n H_{n (\mu - 1) + j} + I_{n (\mu-1) + j,\, n (\mu-1) + j + 1} 
\end{eqnarray}
and $I$ contains the interaction terms of each group with its neighbor group,
\begin{equation}
I = \sum_{\mu=1}^{N_G} I_{\mu n,\mu n + 1}.
\end{equation}
The eigenstates of the group Hamiltonian $H_0$ are products of group eigenstates,
\begin{equation}
\label{prodstate}
H_0 \: \ve a \ket = E_a \: \ve a \ket \enspace \enspace \text{with} \enspace \enspace
\ve a \ket = \prod_{\mu = 1}^{N_G} \ve a_{\mu} \ket ,
\end{equation}
where $\left( \mathcal{H}_{\mu} - I_{\mu n, \mu n + 1} \right) \ve a_{\mu} \ket = E_{\mu} \ve a_{\mu} \ket$.
$E_{\mu}$ is the energy of one subgroup only and $E_a = \sum_{\mu=1}^{N_G} E_{\mu}$.

We assume that the total system is in a thermal state,
\begin{equation}
\label{candens}
\hat \rho = \frac{e^{- \beta H}}{Z} ,
\end{equation}
where $Z$ is the partition sum and $\beta = (k_B T)^{-1}$
the inverse temperature with Boltzmann's constant $k_B$ and temperature $T$.

For our purpose, we need to know the representation of $\hat \rho$ in the basis formed by the
product states $\ve a \ket$. The diagonal elements $\bra a \ve \hat \rho \ve a \ket$ are the
expectation values of $\hat \rho$ in the states $\ve a \ket$,
\begin{equation}
\label{newrho}
\bra a \ve \hat \rho \ve a \ket =
\int_{E_0}^{E_1} w_a (E) \: \frac{e^{- \beta E}}{Z} \: dE ,
\end{equation}
where $w_a (E)$ is the probability density of the occurrence of the energy eigenvalue $E$ in the state
$\ve a \ket$. $E_0$ is the energy of the ground state and $E_1$ the upper limit of the
spectrum. For systems with an energy spectrum that does not have an upper bound, the limit
$E_1 \rightarrow \infty$ should be taken.

One can show that a quantum central limit theorem exists for the present model
\cite{Hartmann2003,Hartmann2004b} and that $w_a (E)$ thus converges to a Gaussian normal distribution
in the limit of infinite number of groups $N_G$,
\begin{equation}
\label{gaussian_dist}
\lim_{N_G \to \infty} w_a (E) = \frac{1}{\sqrt{2 \pi} \Delta_a}
\exp \left(- \frac{\left(E - E_a - \varepsilon_a \right)^2}{2 \, \Delta_a^2} \right),
\end{equation}
where the quantities $\varepsilon_a$ and $\Delta_a$ are defined by 
\begin{eqnarray}
\varepsilon_a & \equiv & \bra a \ve H \ve a \ket - E_a \\
\Delta_a^2 & \equiv & \bra a \ve H^2 \ve a \ket - \bra a \ve H \ve a \ket^2.
\end{eqnarray}
$\varepsilon_a$ is the difference between the energy expectation value of the distribution $w_a (E)$ and the
energy $E_a$, while $\Delta_a^2$ is the variance of the energy $E$ for the distribution $w_a (E)$.
Note that $\varepsilon_a$ has a classical counterpart while $\Delta_a^2$ is purely quantum mechanical.
It appears because the commutator $[H,H_0]$ is nonzero, and the distribution $w_a(E)$ therefore has nonzero
width.

The rigorous proof of eq. (\ref{gaussian_dist}) is given in \cite{Hartmann2003} and based on
the following two assumptions:
The energy of each group $\mathcal{H}_{\mu}$ as defined in eq. (\ref{isogroups}) is bounded, i. e.
\begin{equation}
\label{bounded}
\bra \chi \ve \mathcal{H}_{\mu} \ve \chi \ket \le C
\end{equation}
for all normalized states $\ve \chi \ket$ and some constant $C$, and
\begin{equation}
\label{vacuumfluc}
\bra a \ve H^2 \ve a \ket - \bra a \ve H \ve a \ket^2 \ge N_G \, C'
\end{equation}
for some constant $C' > 0$.

If conditions (\ref{bounded}) and (\ref{vacuumfluc}) are met, eq. (\ref{newrho})
can be computed for $N_G  \gg 1$ \cite{Hartmann2004}:
\begin{equation}
\label{newrho2}
\begin{split}
\bra a \ve \hat \rho \ve a \ket = \frac{1}{2 \, Z} \,
\exp \left(- \beta y_a + \frac{\beta^2 \Delta_a^2}{2} \right) \hspace{2.6cm} \\
\left[\text{erfc} \left( \frac{E_0 - y_a + \beta \Delta_a^2}{\sqrt{2} \, \Delta_a} \right) -
\text{erfc} \left( \frac{E_1 - y_a + \beta \Delta_a^2}{\sqrt{2} \, \Delta_a} \right) \right]
\end{split}
\end{equation}
where $y_a = E_{a} + \varepsilon_a$ and $\text{erfc} (x)$ is the conjugate Gaussian error function \cite{Abramowitz1970}.
The second error function in (\ref{newrho2}) only appears if the energy is bounded and the
integration extends from the energy of the ground state $E_0$ to the upper limit of the spectrum $E_1$.

Note that $y_a$ is a sum of $N_G$ terms and that $\Delta_a$ fulfills eq. (\ref{vacuumfluc}).
The arguments of the conjugate error functions thus grow proportional to $\sqrt{N_G}$ or stronger.
If these arguments divided by $\sqrt{N_G}$ are finite (different from zero),
the asymptotic expansion of the error function \cite{Abramowitz1970} may thus be used for $N_G \gg 1$: $\text{erfc}(x) \rightarrow \exp \left(- x^2 \right) / (\sqrt{\pi} \, x)$ for $x \rightarrow \infty$
and $\text{erfc}(x) \rightarrow 2 + \left(\exp \left(- x^2 \right)\right) / (\sqrt{\pi} \, x)$ for $x \rightarrow - \infty$.
Inserting this approximation into eq. (\ref{newrho2}) and using $E_0 < y_a < E_1$ shows
that the second conjugate error function, which contains the upper limit of the energy spectrum,
can always be neglected compared to the first, which contains the ground state energy.

The off diagonal elements $\bra a \ve \hat \rho \ve b \ket$ vanish for\linebreak 
$\ve E_a - E_b \ve > \Delta_a + \Delta_b$ because the overlap of the two distributions of conditional
probabilities becomes negligible.\linebreak For $\ve E_a - E_b \ve < \Delta_a + \Delta_b$, the transformation
involves an integral over frequencies and thus these terms are significantly smaller than
the entries on the diagonal.

We now test under what conditions the density matrix $\hat \rho$ may be approximated by a product
of canonical density matrices with temperature $\beta_{\text{loc}}$ for each subgroup $\mu = 1, 2, \dots, N_G$.
Since the trace of a matrix is invariant under basis transformations, it is sufficient to verify
the correct energy dependence of the product density matrix.
If we assume periodic boundary conditions,
all reduced density matrices are equal and their product
is of the form $\bra a \ve \hat \rho \ve a \ket \propto \exp(- \beta_{\text{loc}} E_a)$.
We thus have to verify whether the logarithm of rhs of eq. (\ref{newrho2}) is a linear function of the energy $E_a$, 
\begin{equation} \label{log}
\ln \left( \bra a \ve \hat \rho \ve a \ket \right) \approx - \beta_{\text{loc}} \, E_a + c,
\end{equation}
where $\beta_{\text{loc}}$ and $c$ are constants.

We exclude negative temperatures ($\beta > 0$).
Eq. (\ref{log}) can only be true for
\begin{equation} \label{cond_const}
\frac{E_a + \varepsilon_a  - E_0}{\sqrt{N_G} \, \Delta_a} > \beta \frac{\Delta_a^2}{\sqrt{N_G} \, \Delta_a} ,
\end{equation}
as can be seen from eqs. (\ref{newrho2}) and the asymptotic expansion of the error
function. To satisfy (\ref{log}),
$\varepsilon_a$ and $\Delta_a^2$ furthermore have to be of the following form:
\begin{equation}
\label{cond_linear_1} 
- \varepsilon_a + \frac{\beta}{2} \, \Delta_a^2 \approx  c_1 E_a + c_2 
\end{equation}
where $c_1$ and $c_2$ are constants.
Note that $\varepsilon_a$ and $\Delta_a^2$ need not be functions of $E_a$ and therefore in general
cannot be expanded in a Taylor series.

To ensure that the density matrix of each subgroup $\mu$ is approximately canonical, one needs to satisfy
(\ref{cond_linear_1}) for each subgroup $\mu$ separately;
\begin{equation}
\label{cond_linear_2} 
- \frac{\varepsilon_{\mu - 1} + \varepsilon_{\mu}}{2} + \frac{\beta}{4} \,
\left(\Delta_{\mu - 1}^2 + \Delta_{\mu}^2 \right)
+ \frac{\beta}{6} \, \tilde{\Delta}_{\mu}^2 \, \approx \, c_1  \, E_{\mu} + c_2
\end{equation}
where $\varepsilon_{\mu} = \bra a \ve I_{\mu n, \mu n + 1} \ve a \ket$
with $\varepsilon_a = \sum_{\mu=1}^{N_G} \varepsilon_{\mu}$,\linebreak
$\Delta_{\mu}^2 = \bra a \ve \mathcal{H}_{\mu}^2 \ve a \ket -
\bra a \ve \mathcal{H}_{\mu} \ve a \ket^2$ and
$\tilde{\Delta}_{\mu}^2 =
\sum_{\nu = \mu-1}^{\mu+1} \bra a \ve \mathcal{H}_{\nu-1} \mathcal{H}_{\nu} +
\mathcal{H}_{\nu} \mathcal{H}_{\nu-1} \ve a \ket -
2 \bra a \ve \mathcal{H}_{\nu-1} \ve a \ket \bra a \ve \mathcal{H}_{\nu} \ve a \ket$.

Temperature becomes intensive, if the constant $c_1$ vanishes,
\begin{equation} \label{intensivity}
\left| c_1 \right| \ll 1 \enspace \enspace \Rightarrow \enspace \enspace \beta_{\text{loc}} = \beta.
\end{equation}
If this was not the case, temperature would not be intensive, although it might exist locally.

It is sufficient to satisfy conditions (\ref{cond_const}) and (\ref{cond_linear_2}) for an adequate energy
range $E_{\text{min}} \le E_{\mu} \le E_{\text{max}}$ only.
For systems composed of a large number of subsystems, the density of states is
typically a rapidly growing function of energy \cite{Tolman1967}. If the total system is in a
thermal state, occupation probabilities decay exponentially with energy. The product of these two
functions is thus sharply peaked at the expectation value of the energy $\overline{E}$ of the total
system $\overline{E} + E_0 = $Tr$(H \hat \rho)$, with $E_0$ being the ground state energy.
Therefore a pertinent and ``safe'' choice for $E_{\text{min}}$ and $E_{\text{max}}$ is
\begin{equation} \label{e_range}
\begin{array}{rcl}
E_{\text{min}} & = & \text{max}
\left( \left[E_{\mu}\right]_{\text{min}} \, , \,
\frac{1}{\alpha} \frac{\overline{E}}{N_G} + \frac{E_0}{N_G} \right)\\
E_{\text{max}} & = & \text{min}
\left( \left[E_{\mu}\right]_{\text{max}} \, , \, \alpha
\frac{\overline{E}}{N_G} + \frac{E_0}{N_G} \right)
\end{array}
\end{equation}
where $\alpha \gg 1$ and $\overline{E}$ will in general depend on the global temperature.
In eq. (\ref{e_range}), $\left[E_{\mu}\right]_{\text{min}}$ and
$\left[E_{\mu}\right]_{\text{max}}$ denote
the minimal and maximal values $E_{\mu}$ can take on.

For a model obeying eqs. (\ref{bounded}) and (\ref{vacuumfluc}), the two conditions
(\ref{cond_const}) and (\ref{cond_linear_2}), which constitute the general result of
this letter, must both be satisfied. These fundamental criteria will now be applied to
a concrete example.

We now consider the Heisenberg-model of a spin chain in a transverse field .
The Hamiltonian reads \cite{vanVleck1945,Manousakis1991}
\begin{equation} \label{ising_ham}
H_i = B \, \sigma_i^z \enspace \, ; \enspace \enspace
I_{i, i+1} = J \left(
\sigma_i^x \sigma_{i+1}^x + \sigma_i^y \sigma_{i+1}^y + \sigma_i^z \sigma_{i+1}^z \right)
\end{equation}
where $\sigma_i^x, \sigma_i^y$ and $\sigma_i^z$ are the Pauli matrices. $B$ is the
magnetic field and $J$ the coupling parameter. We will always assume $B > 0$.

We partition the chain into $N_G$ groups of $n$ adjacent spins, as considered above,
and numerically analyze eqs. (\ref{cond_const}) and (\ref{cond_linear_2}) by
exact diagonalization of the groups. In doing so, we apply the approximation
$\tilde{\Delta}_{\mu}^2 \ll \Delta_{\mu}^2$.
Since $\tilde{\Delta}_{\mu}^2 =
\sum_{\nu = \mu-1}^{\mu+1} \bra a \ve I_{\nu-1} I_{\nu} +
I_{\nu} I_{\nu-1} \ve a \ket -
2 \bra a \ve I_{\nu-1} \ve a \ket \bra a \ve I_{\nu} \ve a \ket$,
only $\sigma^z \otimes \sigma^z$ terms contribute and, among those, only terms which
are products of one diagonal element and one off-diagonal element of the two $\sigma^z$.
$\Delta_{\mu}^2$ thus contains much more terms, which,
in addition, are all positive.

The conditions for the central limit theorem are met for almost all states $\ve a \ket$
of the present model:
Condition (\ref{bounded}) is fulfilled because the Hamiltonian of a single spin has
finite dimension. The numerics shows that for almost all states, $\Delta_{\mu}^2$
is a finite positive number, which implies that condition (\ref{vacuumfluc}) is satisfied.

For single spins, a local temperature can always be defined by assigning a
Boltzmann factor to the occupation probabilities of the upper and lower level.
This local temperature, however, is not equal to the global one.

For groups of more than one spin,
neither $\varepsilon_{\mu}$ nor $\Delta_{\mu}^2$ can be approximated by linear functions
of $E_{\mu}$. Therefore local temperature can only exists if the rhs of eq.
(\ref{cond_linear_2}) is a constant. As a direct consequence, temperature is intensive,
$\beta_{\text{loc}} = \beta$.

Eq. (\ref{cond_const}) can be checked directly. To check eq. (\ref{cond_linear_2}),
we need to quantify what ``approximately constant'' means. We thus use
\begin{equation}
\label{min_max_linear}
\begin{split}
\beta \, \frac{\left[ \Delta_{\mu}^2 \right]_{\text{max}} - \left[ \Delta_{\mu}^2 \right]_{\text{min}}}{2} +
\left[ \varepsilon_{\mu} \right]_{\text{max}} - \left[ \varepsilon_{\mu} \right]_{\text{min}} \ll\\
\ll \left[ E_{\mu} \right]_{\text{max}} - \left[ E_{\mu} \right]_{\text{min}} ,
\end{split}
\end{equation}
where $[ x ]_{\text{max}}$ and $[ x ]_{\text{min}}$ denote the maximal and minimal value $x$ takes on in
all states $\ve a \ket$. Condition (\ref{min_max_linear}) implies, that (\ref{cond_linear_2})
holds.

\begin{figure}[t]
\centering
\psfrag{0.1}{\small \hspace{-0.1cm} \raisebox{-0.1cm}{$1$}}
\psfrag{1.1}{\small \hspace{-0.15cm} \raisebox{-0.1cm}{$10^{1}$}}
\psfrag{2.1}{\small \hspace{-0.15cm} \raisebox{-0.1cm}{$10^{2}$}}
\psfrag{3.1}{\small \hspace{-0.15cm} \raisebox{-0.1cm}{$10^{3}$}}
\psfrag{1}{\small \hspace{-0.3cm} \raisebox{-0.0cm}{$1$}}
\psfrag{2}{\small \hspace{-0.3cm} \raisebox{-0.0cm}{$2$}}
\psfrag{3}{\small \hspace{-0.3cm} \raisebox{-0.0cm}{$3$}}
\psfrag{4}{\small \hspace{-0.3cm} \raisebox{-0.0cm}{$4$}}
\psfrag{5}{\small \hspace{-0.3cm} \raisebox{-0.0cm}{$5$}}
\psfrag{6}{\small \hspace{-0.3cm} \raisebox{-0.0cm}{$6$}}
\psfrag{7}{\small \hspace{-0.3cm} \raisebox{-0.0cm}{$7$}}
\psfrag{n}{\hspace{-0.2cm} $n_{\text{min}}$}
\psfrag{T}{\hspace{-0.0cm} \raisebox{-0.0cm}{$T / B$}}
\includegraphics[width=8cm]{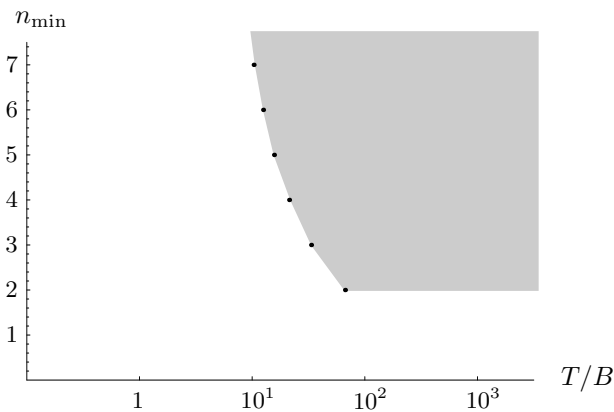}
\caption{$n_{\text{min}}$ as a function of the temperature $T$ for $J = B > 0$.
$T$ is given in units of $B$, $k_B = 1$ and $\alpha = 10$.
$\alpha$ is defined in eq. (\ref{e_range}).
Local temperature exists in the shaded region.
It also exists for single spins but is never intensive there.}
\label{logmittela}
\end{figure}

By numerical evaluation of eqs. (\ref{cond_const}) and (\ref{min_max_linear}), one
can calculate a minimal group size $n_{\text{min}} > 2$ for each temperature $T$, where
we consider (\ref{min_max_linear}) to be satisfied if the lhs is 100 times smaller than the rhs.
This corresponds to a tolerable deviation from the canonical distribution of $1 \%$.
Figure \ref{logmittela} shows $n_{\text{min}}$ as a function of $T$ for
antiferromagnetic coupling with strength $J = B$. Local temperature can exist in the shaded
region. As mentioned above, local temperature can always be defined for single spins.
For the present coupling strength, however, it is not intensive unless $T \rightarrow \infty$.
%
%
%
%

\begin{figure}[t]
\centering
\psfrag{0.1}{\small \hspace{-0.1cm} \raisebox{-0.1cm}{$1$}}
\psfrag{1.1}{\small \hspace{-0.15cm} \raisebox{-0.1cm}{$10^{1}$}}
\psfrag{2.1}{\small \hspace{-0.15cm} \raisebox{-0.1cm}{$10^{2}$}}
\psfrag{3.1}{\small \hspace{-0.15cm} \raisebox{-0.1cm}{$10^{3}$}}
\psfrag{1}{\small \hspace{-0.3cm} \raisebox{-0.0cm}{$1$}}
\psfrag{2}{\small \hspace{-0.3cm} \raisebox{-0.0cm}{$2$}}
\psfrag{3}{\small \hspace{-0.3cm} \raisebox{-0.0cm}{$3$}}
\psfrag{4}{\small \hspace{-0.3cm} \raisebox{-0.0cm}{$4$}}
\psfrag{5}{\small \hspace{-0.3cm} \raisebox{-0.0cm}{$5$}}
\psfrag{6}{\small \hspace{-0.3cm} \raisebox{-0.0cm}{$6$}}
\psfrag{7}{\small \hspace{-0.3cm} \raisebox{-0.0cm}{$7$}}
\psfrag{n}{\hspace{-0.2cm} $n_{\text{min}}$}
\psfrag{T}{\hspace{-0.0cm} \raisebox{-0.0cm}{$T / B$}}
\includegraphics[width=8cm]{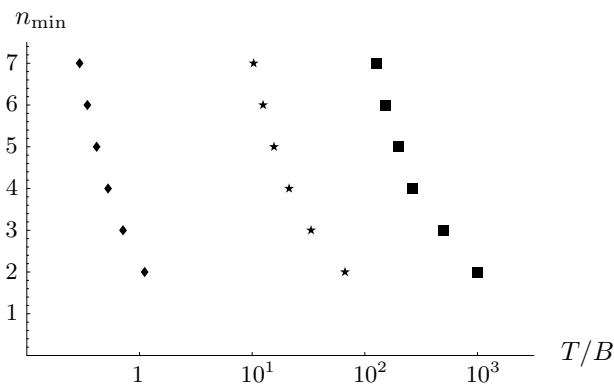}
\caption{$n_{\text{min}}$ $n_{\text{min}}$ as a function of the temperature $T$ for
$J = 0.1 B$ (diamonds), $J = B$ (stars) and $J = 10 B$ (squares), $J > 0$ and $B > 0$.
$T$ is given in units of $B$, $k_B = 1$ and $\alpha = 10$.
$\alpha$ is defined in eq. (\ref{e_range}).
For each value of $J$, local temperature exists for $n > n_{\text{min}}$.
For single spins, local temperature always exists, but is only intensive for $J = 0.1 B$
and $T > 2.5 B$.}
\label{vergl}
\end{figure}

Figure \ref{vergl} shows $n_{\text{min}}$ as a function of the temperature $T$ for
antiferromagnetic couplings of different strength.
Diamonds correspond to $J = 0.1 B$, stars to $J = 1 B$ and squares to $J = 10 B$.
The regions, where local temperatures exist are as indicated in figure \ref{logmittela}.
For single spins, local temperature always exists, but is only intensive for $J = 0.1 B$
and $T > 2.5 B$
%
%
%
%

In both plots, only the one of eqs. (\ref{cond_const}) and (\ref{cond_linear_2}),
which sets the stronger bound enters. For the present parameter range, this is exclusively
eq. (\ref{cond_linear_2}).
The results for the ferromagnetic case are approximately the same and are therfore not given
separately here.

In summary, we have considered a homogeneous chain of quantum systems with nearest neighbor interactions.
We have partitioned the chain into identical groups of $n$ adjoining subsystems each.
Taking the number of such groups to be very large and assuming the total
system to be in a thermal state with temperature $T$ we have found conditions
(eqs. (\ref{cond_const}) and (\ref{cond_linear_2})), which ensure that each group is approximately
in a thermal state. Furthermore, we have determined when the isolated groups have the same temperature $T$,
i.e. when temperature is intensive.

The result shows that, in the quantum regime,
these conditions depend on the temperature $T$, contrary to the classical case.
The characteristics of the temperature dependence are determined by the width $\Delta_a$ of the distribution
of the total energy eigenvalues in a product state and its dependence on the group energies $E_a$.
The low temperature behavior, in particular, is related to the fact that $\Delta_a$ has a nonzero
minimal value. This fact does not only appear in spin chains 
but is a general feature of quantum systems composed of interacting particles or subsystems.
The commutator $[H,H_0]$ is nonzero and the ground state of the total
system is energetically lower than the lowest product state, therefore $\Delta_a$ is nonzero, even
at zero temperature \cite{Jordan2003}.

For the models we consider here, the off diagonal elements of the density operator in
the product basis, $\bra a \ve \hat \rho \ve b \ket$ ($a \not= b$), are
significantly smaller than the diagonal ones, $\bra a \ve \hat \rho \ve a \ket$.
Our general result, conditions (\ref{cond_const}) and (\ref{cond_linear_2}), thus
states that the density matrix $\hat \rho$ ``approximately'' factorizes with respect to
the considered partition. This implies that the state $\hat \rho$ is not entangled
with respect to this partition, at least within the chosen accuracy.
It would therefore be interesting to see how our result relates to
the scaling of entanglement in many particle systems \cite{Vidal2003}.

Unfortunately, our approach only applies to nonzero temperatures. The underlying central limit theorem
\cite{Hartmann2003,Hartmann2004b} is about the weak convergence of the distribution of energy eigenvalues.
Weak convergence means that only integrals over energy intervals of nonzero length do converge.
We thus cannot make statements about a system in its ground state let alone about
the entanglement in that state.

We have then applied the general method to several types of Heisenberg spin chains.
For concrete models, the conditions (\ref{cond_const}) and (\ref{cond_linear_2}) determine a 
minimal group size and thus a minimal length scale
on which temperature may be defined according to the temperature concept we adopt.
Grains of size below this length scale are no more in a thermal state. Thus temperature measurements
with a higher resolution should no longer be interpreted in a standard way.

The length scales, calculated in this paper, should also constrain the way one can meaningfully
define temperature profiles in non-equilibrium scenarios \cite{Michel2003}.

\end{document}